\documentclass[12pt]{article}
\def\ee{\end{equation}}
\def\be{\begin{equation}}
\def\ba{\begin{eqnarray}}
\def\ea{\end{eqnarray}}
\begin{document}
\baselineskip 24pt
%
%
\title{Additive entropy underlying the general composable
entropy prescribed by thermodynamic meta-equilibrium}
\author{{Ramandeep S. Johal}
\footnote{e-mail: rjohal@theory.phy.tu-dresden.de}\\
{Institute for Theoretical Physics,}\\
{Technical University, D-01062 Dresden, Germany }}
\maketitle
\begin{abstract}
We consider the meta-equilibrium state of a composite  system 
made up of independent subsystems satisfying the additive form of
external constraints, as recently discussed by Abe [Phys. Rev. E
{\bf 63}, 061105 (2001)]. We derive the additive entropy $S$ underlying
a composable entropy $\tilde{S}$ by identifying the common 
intensive variable. The simplest
form of composable entropy satisfies Tsallis-type nonadditivity and
the most general composable form is interpreted as a
monotonically increasing funtion $H$ of this simplest form. 
 This is consistent with the observation that the
meta-equilibrium can be equivalently described by the
maximum of either $H[\tilde{S}]$ or $\tilde{S}$ and the intensive 
variable is same in both cases.

\noindent {\it PACS Number(s):  05.20.-y, 05.70.-a, 05.90.+m}
\end{abstract}
\newpage
Generalized entropic functionals which may
yield consistent formulations of thermodynamics \cite{abepre01}
and statistical mechanics \cite{kaniadakis02} has attracted
attention in recent years. Generalization of the standard
entropy can also be a useful tool to describe effects of finiteness
in physical systems \cite{gorbankarlin03}. The standard thermodynamic
formalism is based on Boltzmann-Gibbs-Shannon entropy,
which is additive. A particular generalization of the standard
formalism that has been  much studied  recently is popularly
called Tsallis statistics \cite{tsallis88}.
This formalism seems to provide an effective description
of the meta-equilibrium states of certain complex systems.
It is based on a generalized
entropy that is nonadditive even when the subsystems forming
the composite system are statistically independent.
Tsallis-type nonadditivity may be considered
as the simplest case of the most general nonadditivity rule for composable
entropy, consistent with the existence of thermodynamic
equilibrium \cite{abepre01,rjohalpp01}. This so called equilibrium
within the Tsallis approach is actually a meta-equilibrium, which is
described by a generalized zeroth law.

On the other hand, it has been suggested that the thermodynamic framework
based on Tsallis entropy can be mapped to the framework based on additive (extensive)
entropy \cite{abemartinez01}-\cite{toral03}. Essentially, it means that a
combination of an additive entropy
and an appropriate intensive variable preserves the  standard thermodynamic
relations  as well as statistical fluctuations.

In this letter, we derive the additive entropy underlying
a composable entropy $\tilde{S}$ by  identifying the intensive
variable common to subsystems in meta-equilibrium. 
The case of the most general form for composable entropy  as
treated in \cite{abepre01} is interpreted as a
monotonically increasing function $H$ of the  simplest composable entropy. 
 This is consistent with the fact that the intensive variable should be  
same in the meta-equilibrium state  corresponding to the maximum of
either $H[\tilde{S}]$ or $\tilde{S}$.  

Tsallis entropy is defined as
\be
S^{(T)}_q = \frac{ \sum_{i=1}^W p_i^q -1}{1-q},
\label{st}\ee 
where $p_i$ is the probability distribution set characterising 
the discrete microstates of the system and labelled $i=1,...,W$. 
The transformation relating additive and nonadditive 
entropies is given by
\be
{S}^{(R)}_q = {1\over (1-q)} {\rm ln} [1 + (1-q){S}^{(T)}_{q}].
\label{ne}
\ee
The additive entropy so obtained is well known as Renyi entropy 
\cite{renyibook}.

For better understanding of this transformation,
we first derive it directly by applying the generalized zeroth law
for meta-equilibrium within Tsallis' framework.
In line with earlier works \cite{abepre01}, we impose additive mean values 
as constraints. Thus we restrict to a special class of nonextensive
models in which the entropy is non-additive, but the energy and other 
external constraints are additive. Furthermore, it is understood that 
the zeroth law in the standard Boltzmann-Gibbs formalism
identifies an intensive variable common to the systems
in mutual thermodynamic equilibrium. The generalized zeroth
law serves a similar purpose within Tsallis approach.

Consider the meta-equilibrium between two systems $A$ and $B$,
such  that the maximum of the
nonadditive entropy $\tilde{S}_{q}$ holds for the composite system
$(A+B)$
\be
\tilde{S}_q(A+B) = \tilde{S}_q(A) + \tilde{S}_q(B) + 
(1-q)\tilde{S}_q(A)\tilde{S}_q(B),
\label{nad}
\ee
under the constraints which are fixed values of $n$ number of
additive quantities 
\be
E_k(A+B) = E_k(A) + E_k(B), \qquad  k=1,...,n.
\label{extcs}
\ee
As in standard thermodynamics, we define the meta-equilibrium
entropy for the nonadditive case to be
$k_{\rm B}\tilde{S}_{q}( \{E_k |k=1,...,n \} )$, where
$k_{\rm B}$ is the Boltzmann's constant. Apparently, 
$\tilde{S}_{q}$ is the 
explicit entropy function in terms of the set of given additive
constraints $\{E_k |{k=1,...,n} \}$ and is obtained from the optimum
distribution $\{p_{i} |{i=1,...,W} \}$ following from the maximisation of
$\tilde{S}_q (\{p_i\})$ \cite{tmp98}.  

Now making the variations $\delta \tilde{S}_{q} (A+B)$
and $\delta E_k(A+B)$  for ${k=1,...,n}$ vanish for the meta-equilibrium 
state, we obtain 
\be
[1 + (1-q)\tilde{S}_{q} (B) ] {\partial \tilde{S}_{q}(A) \over \partial E_k(A)}
= [1 + (1-q)\tilde{S}_{q} (A) ] {\partial \tilde{S}_{q}(B) \over \partial E_k(B)}.
 \label{vars}
\ee
On rearranging (\ref{vars}), we can achieve separation of variables as
\be
{1 \over [1 + (1-q)\tilde{S}_{q} (A) ]} {\partial \tilde{S}_{q}(A)  
\over \partial E_k(A)} =
{1 \over [1 + (1-q)\tilde{S}_{q} (B) ]} {\partial \tilde{S}_{q}(B) 
\over \partial E_k(B)} = \eta_k.
\label{rear}
\ee
The parameter $\eta_k$,  by definition, is common to both
the subsystems $A$ and $B$, and is identified as the intensive variable.
Dropping the index $A$ or $B$, for each subsystem the following 
equation holds for all values of $k$
\be
{1 \over [1 + (1-q)\tilde{S}_{q} ]}{\partial \over \partial E_k}
\tilde{S}_{q} (E_1,...,E_k,...,E_n)  = \eta_k.
\label{deq}
\ee
Note that ${\partial \tilde{S}_{q} \over \partial E_k}$
defines the Lagrange multiplier associated with the 
maximum entropy method and so according to the above
equation, intensive variables and the corresponding 
Lagrange multipliers are not identical, except when
the entropy is additive.

The relation with an additive entropy can be established if we 
assume  that the intensive variable
$\eta_k$ occuring above is the same as defined for an 
additive entropy $S(E_1,...,E_k,...,E_n)$ and is given by
\be
\eta_k = {\partial{S} \over \partial E_k}.
\label{intensive}
\ee
We remark here that the choice of the form of additive entropy
is arbitrary here; one is free to use either BGS entropy or 
Renyi entropy.
From (\ref{deq}) and (\ref{intensive}), this implies that
the nonadditive and the additive entropies are related in the
following way 
\be
{1\over (1-q)} {\rm ln}[1 + (1-q)\tilde{S}_q]  = S + c,
\label{int}
\ee
where the constant $c$ has to be independent of $E_k$. 
Thus  the transformation (\ref{ne})
of a nonadditive entropy to an additive one is recovered   for the 
particular choice of $c=0$. 

Next, we explore this relation between nonadditive and
additive  entropies from the viewpoint of composable entropies
as prescribed by the existence of meta-equilibrium.
To proceed further, we recall the notion of composability of entropy
\cite{tsallis88,hotta99}.
An arbitrary entropic form $\tilde{S}$ is  defined to be composable, if 
the total entropy $\tilde{S}(A, B)$ for the composite system can be
written as
$\tilde{S}(A, B) = f[\tilde{S}(A),\tilde{S}(B)]$,
where $f[\cdot ]$ is a certain bivariate function 
of the $C^2$ class and is symmetric in its arguments,
$ f[\tilde{S}(A),\tilde{S}(B)] = f[\tilde{S}(B),\tilde{S}(A)]$.

Suppose the maximum of the general entropy determines a kind of
meta-equilibrium between subsystems $A$ and $B$, under the given additive
constraints (\ref{extcs}). By equating the variations of the total
entropy and total value of the constraint quantity to zero, 
we get
\be
{\partial \tilde{S}(A, B) \over
\partial \tilde{S}(A) } {\partial \tilde{S}(A)
\over \partial E_k(A)} =
{\partial \tilde{S}(A, B) \over
\partial \tilde{S}(B) } {\partial \tilde{S}(B)
\over \partial E_k(B)}.  
\label{var0}
\ee
To establish an intensive variable common to the two subsystems,
a general case can be written in the following form \cite{abepre01}
\ba
{\partial \tilde{S}(A, B) \over
\partial \tilde{S}(A) } &=&{1\over \lambda} G[\tilde{S}(A, B)]
 {d h[\tilde{S}(A)] \over
d \tilde{S}(A)} h[\tilde{S}(B)],\label{factgh} \\
{\partial \tilde{S}(A, B) \over
\partial \tilde{S}(B) } &=& {1\over \lambda}G[\tilde{S}(A, B)] 
h[\tilde{S}(A)] {d h[\tilde{S}(B)] \over d \tilde{S}(B)},
\label{facthg}
\ea
where $h[\cdot ]$ is some differentiable function; $G[\cdot ]$
is also an arbitrary function and $\lambda$ is a constant.
Therefore,
using (\ref{factgh}) and (\ref{facthg}) in (\ref{var0}) and rearranging,
 an intensive variable common to systems $A$ and $B$ can be defined as
\be
{1\over \lambda}  {1\over {h[\tilde{S}(A)]}}{d h[\tilde{S}(A)] \over
d \tilde{S}(A)} {\partial \tilde{S}(A)
\over \partial E_k(A)} =
{1\over \lambda} {1\over {h[\tilde{S}(B)]}}{d h[\tilde{S}(B)] \over
d \tilde{S}(B)} {\partial \tilde{S}(B)
\over \partial E_k(B)} = \eta_k.
\label{intr}
\ee
Again, ${\partial \tilde{S} \over \partial E_k}=\tilde{\eta}_{k}$
is the Lagrange multiplier  associated with system $A$ or $B$.
Thus (\ref{intr}) defines the relation between the Lagrange multiplier
and the intensive variable for a general composable entropy.

We emphasize that in the present approach the 
intensive variable associated with a general composable entropy is {\it 
independent of the function $G$}, which by definition, is not
factorisable into contributions from systems $A$ and $B$.

For any subsystem ($A$ or $B$), we can rewrite (\ref{intr}) as
\be
{1\over \lambda h[\tilde{S}] } {\partial h[\tilde{S}]
\over \partial E_k} = \eta_k.
\label{partialhek}
\ee
Again invoking the relation (\ref{intensive}), we note that the 
function $h$ is related to the additive entropy $S$ as
 $h[\tilde{S}] = {\exp}(\lambda \{ S  +c \} )$, 
where  $c$ has been identified before. Setting $\tilde{S} =0$ 
for $S=0$, we obtain
\be
h[0] = {\exp}(\lambda c).
\label{formh0}
\ee
Note that for $\lambda\to 0$, we have $h[0]\to 1$,
a property assumed in \cite{abepre01}. Alternately,  
if the constant is set as $c=0$, then $h[0]$
is unity for all values of $\lambda$. In general, we write
\be
h[\tilde{S}] = h[0]{\rm exp}(\lambda S).
\label{exph}
\ee 
{\it This form of the function $h$ is solely determined by the 
requirement of the
existence of an intensive variable associated with an 
additive entropy $S$.} It is thus intrinsically independent of the form of the
$G[\cdot ]$ function.

We remark here that Tsallis type nonadditivity of 
entropy is obtained as a special
case of the above analysis, if we put $G$ as a constant
equal to unity, and identify for each  {\it subsystem}
\be
h[\tilde{S}] =  1+ \lambda \tilde{S},
\label{hsts}
\ee
which gives $h[0]=1$.
Then equation (\ref{intr}) is identical to (\ref{rear}) for $\lambda = (1-q)$.
On combining (\ref{exph}) and (\ref{hsts}), we have 
\be
\tilde{S} = {\exp(\lambda S) -1 \over \lambda},
\ee
which satisfies
\be
\tilde{S}(A+B) = \tilde{S}(A) + \tilde{S}(B) + 
\lambda \tilde{S}(A)\tilde{S}(B).
\label{nad2}
\ee
Next, we ascertain the relation between the  
additive entropy $S$ and the composable entropy $\tilde{S}$
for a general function $G$. Following \cite{abepre01}, it
is convenient to set  
\be
G[\tilde{S}(A, B) ] = \left( {d H[\tilde{S}(A, B) ] \over
                        d \tilde{S}(A, B) }\right)^{-1}.
                        \label{defineg}
                        \ee
This allows to obtain the composability rule for entropy in terms of
the function $H$.
The case of Tsallis type nonadditivity correpsonds in this
notation to $H$ as an identity function.
For concreteness, let us assume that the function $G[\cdot ]$
is positive, implying that $H[\tilde{S} ]$ is a monotonically
increasing (differentiable) function of $\tilde{S}$. 
We give an interpretation of this assumption later on.
So using (\ref{defineg}), the relations (\ref{factgh}) and (\ref{facthg}) 
are rewritten as
\ba
{d H[\tilde{S}(A, B) ] \over d \tilde{S}(A, B) }
{\partial \tilde{S}(A, B) \over
\partial \tilde{S}(A) } &=&{1\over \lambda}
 {d h[\tilde{S}(A)] \over
d \tilde{S}(A)} h[\tilde{S}(B)],\label{factgh2} \\
{d H[\tilde{S}(A, B) ] \over d \tilde{S}(A, B) } 
{\partial \tilde{S}(A, B) \over
\partial \tilde{S}(B) } &=& {1\over \lambda} 
h[\tilde{S}(A)] {d h[\tilde{S}(B)] \over d \tilde{S}(B)},
\label{facthg2}
\ea
which further imply
\ba
{\partial H[\tilde{S}(A, B) ] \over \partial \tilde{S}(A) }
 &=&{1\over \lambda} {d h[\tilde{S}(A)] \over
d \tilde{S}(A)} h[\tilde{S}(B)],\label{factgh3} \\
{\partial H[\tilde{S}(A, B) ] \over \partial \tilde{S}(B) } 
 &=& {1\over \lambda} 
h[\tilde{S}(A)] {d h[\tilde{S}(B)] \over d \tilde{S}(B)}.
\label{facthg3}
\ea
On integrating (\ref{factgh3}) or (\ref{facthg3}), we obtain   
the composability rule for entropy  as
\be
H[\tilde{S}(A, B) ] =
{1 \over \lambda} h[\tilde{S}(A)] h[\tilde{S}(B)] + {\rm constant}.
\label{solH}
\ee
The constant is determined by the requirement that 
$\tilde{S}(A, B)=0$ for $\tilde{S}(A)= \tilde{S}(B)=0$.
Finally on using (\ref{exph}), we obtain
\be
H[\tilde{S}(A, B) ] =
{h^2[0]\over \lambda}\big\{ \rm{exp}[\lambda \{S(A)+S(B) \}] -1 \big\} + H[0].
\ee
Using the fact that for an ordered system, say $B$,
$S(B)=0$, which implies  $\tilde{S}(A, B)=\tilde{S}(A)$,
we obtain the transformation between an additive entropy
and the general composable entropy for a subsystem ($A$ or $B$)
as follows
\be
S = {1\over \lambda} {\rm ln}\;\left[
1+ {\lambda\over h^2[0]} ( H[\tilde{S}] - H[0] )\right].
\label{sexth}
\ee
This relation may be taken as the most general transformation
connecting additive and composable nonadditive entropies compatible with the
generalized zeroth law.  
Note that it is not implied that any composable entropy can be mapped 
to an additive entropy. As we argue below, the identification of
an intensive variable which is independent of the function $G$ or $H$,
seems to imply that the composable entropy is either $\tilde{S}$
which satisfies Tsallis-type nonadditivity, or it is a monotonically
increasing function $H$ of $\tilde{S}$.

To illustrate this conclusion, suppose that
the meta-equilibrium is described by the maximum of
$H[\tilde{S}(A, B)]$ under the additive constraints (\ref{extcs}).
Then making the variation $\delta H$ vanish under 
$\delta E_k(A) = -\delta E_k(B)$,
the condition of equilibrium implies
\be
{\partial H[\tilde{S}(A, B)] \over
\partial \tilde{S}(A) } {\partial \tilde{S}(A)
\over \partial E_k(A)} =
{\partial H[\tilde{S}(A, B)] \over
\partial \tilde{S}(B) } {\partial \tilde{S}(B)
\over \partial E_k(B)}.
\label{varH}
\ee 
This can be rewritten as 
\be
{d H[\tilde{S}(A, B)] \over
d \tilde{S}(A,B) }  {\partial \tilde{S}(A, B) \over
\partial \tilde{S}(A) }  {\partial \tilde{S}(A)
\over \partial E_k(A)} =
{d H[\tilde{S}(A, B)] \over
d \tilde{S}(A, B) }  {\partial \tilde{S}(A, B) \over
\partial \tilde{S}(B) }  {\partial \tilde{S}(B)
\over \partial E_k(B)},
\label{varH2}
\ee 
Now assuming the conditions (\ref{factgh}), (\ref{facthg})
and on using (\ref{defineg}) we obtain the same relation as
(\ref{intr}). Thus the intensive variable corresponding
to the maximisation of $\tilde{S}(A, B)$ is identical to that
for the maximisation of $H[\tilde{S}(A, B)]$, which is
in accordance with the fact that $H$ has been assumed to
be a monotonic function of $\tilde{S}$ and thus the 
equilibrium condition obtained from each is
identical under similar constraints. Again note that
for $H$ to be identical function, the above equations
go neatly to the case of the simplest composable entropy.
 
It may be pointed out that although the intensive variable
is the same for the maximisation of either $\tilde{S}$ or $H[\tilde{S}]$,
yet the relation between intensive variable $\eta_k$ and the corresponding
Lagrange multiplier $\tilde{\eta}_{k}$ depends on the specific entropy chosen
for maximisation. For the case of Tsallis entropy, 
the  intensive variable is given by \cite{abemartinez01,vives02}
\be
\eta_k = {\tilde{\eta}_k \over \{1 + (1-q){S_q}^{({\rm T})} \} }.
\label{lagin}
\ee
To obtain the corresponding relation for the case of most general
composable entropy, we can apply the partial derivative with respect to
$E_k$ to Eq. (\ref{sexth}) and obtain
\be
\eta_k = {\tilde{\eta}_{k}^{(H)} \over \{ h^2[0]+ \lambda ( H[\tilde{S}] -H[0] ) \} }
        {d H[\tilde{S} ] \over d \tilde{S} }.
        \label{genlmt}
        \ee
Summarising, it has been shown in literature that an additive
entropy underlies the Tsallis entropy, which alongwith an 
intensive variable, preserves the standard
thermodynamic structure. In this letter, we have looked at the relation
between nonadditive and additive entropies from the
viewpoint of the generalised zeroth law and the notion of
composable entropies. 
We have shown  the mapping between  Tsallis-type nonadditive 
entropy (which is the simplest  composable form consistent with
meta-equilibrium) and an additive entropy 
by identifying the intensive variable common to two 
systems in meta-equilibrium. The abovementioned
mapping becomes possible if this intensive variable
is also the one associated with an additive entropy. 
Further, we have argued that if the meta-equilibrium is 
alternately described
by the maximum of  the most general composable entropy $H$,
then $H$ can be interpreted as a monotonically increasing
function of the Tsallis-type nonadditive entropy. 
The fact that the meta-equilibrium state is expected to be physically 
the same in either maximisation problem, is consistent with the
observation that the intensive variable is same in
both the cases.   

Tha author is grateful to Professor Gerhard Soff for encouragement
and support, and to Professor Sumiyoshi Abe for  many useful 
discussions.

\end{document}